\documentclass [prb,superscriptaddress, showpacs, twocolumn]{revtex4}
\usepackage{graphicx}
\usepackage{dcolumn}

\begin{document}

\title{Spontaneous toroidic effects in Ba$_2$CoGe$_2$O$_7$}

\author{Pierre Toledano}
\affiliation{Laboratory of Physics of Complex Systems, University of Picardie, 33 rue Saint-Leu, 80000 Amiens, France}
\author{Dmitry D. Khalyavin}
\affiliation{ISIS facility, STFC Rutherford Appleton Laboratory, Chilton, Didcot, Oxfordshire, OX11-0QX,United Kingdom}
\author{Laurent C. Chapon}
\affiliation{ISIS facility, STFC Rutherford Appleton Laboratory, Chilton, Didcot, Oxfordshire, OX11-0QX,United Kingdom}

\date{\today}

\begin{abstract}
The unusual magnetoelectric effects observed in the multiferroic phase arising below $T_N$=6.7K in Ba$_2$CoGe$_2$O$_7$ (BCG) are related to  the spontaneous toroidal moment existing in this compound. The transition to the multiferroic state, which involves spontaneous magnetization, polarization and toroidal moment gives rise to spontaneous toroidic effects. These effects produce specific contributions to the spontaneous polarization and magnetization under applied magnetic or electric fields which provide indirect indications of the existence and role of the toroidal moment in multiferroic materials. The toroidic contribution to the electric polarization in BCG is shown to result from single-ion effects.
\end{abstract}


\maketitle

\begin{figure*}[t]
\includegraphics[scale=0.76]{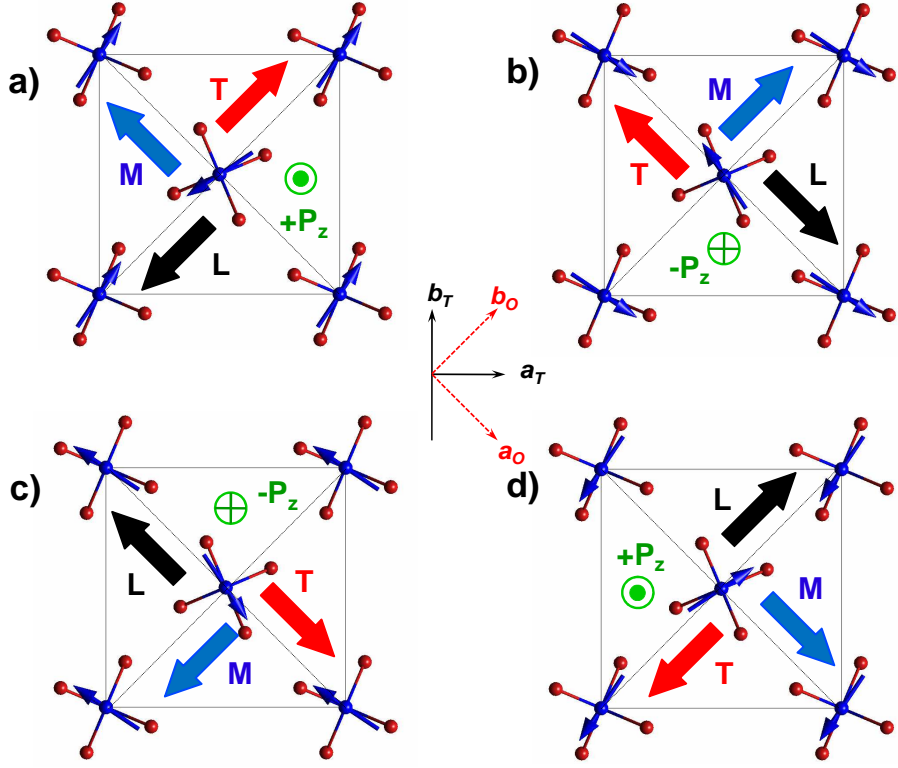}
\caption{(Color online) Respective orientations of the magnetization ($\vec{M}$), antiferromagnetic vector ($\vec{L}$), toroidal moment ($\vec{T}$) and polarization ($\vec{P}$) in the four multiferroic domains of BCG. Dark blue thin arrows represent the orientation of the spins $\vec{s_1}$ and $\vec{s_2}$ located in positions (0,0,0) and (0.5,0,5,0) in the tetragonal paramagnetic structure. The central inset shows the tetragonal and orthorhombic settings used in the text.}
\label{fig:F1}
\end{figure*}

\begin{figure}[t]
\includegraphics[scale=1.0]{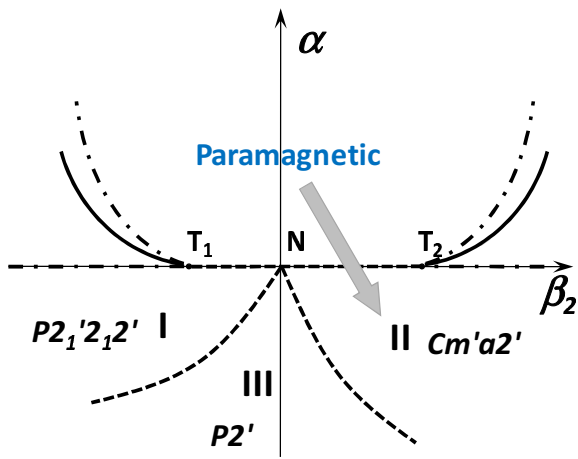}
\caption{(Color online) Theoretical phase diagram associated with the free-energy $F$ given by Eg. (\ref{eq:F2}). Solid and hatched curves are first and second-order transitions lines. Hatched-dotted curves are limits of stability lines. T$_1$ and T$_2$ are tricritical points. N is a four-phase point. The arrow represents the thermodynamic path followed in BCG.}
\label{fig:F2}
\end{figure}

\indent The resurgence of interest in multiferroic materials has prompted discussion of the relevance of the concept of magnetic toroidal moment for clarifying the macroscopic and microscopic properties of these systems. \cite{ISI:1,ISI:2,ISI:3} The existence of a macroscopic moment asymmetric under both time reversal and space inversion long remained elusive\cite{ISI:4,ISI:5} until the observation of the independent coexistence of ferrotoroidic and antiferromagnetic domains in LiCoPO$_4$.\cite{ISI:6} This result provides a motivation for investigating toroidic effects in the ferroelectric phases of magnetic multiferroic materials, in which the space-asymmetric electric polarization is induced by a time-asymmetric and space-asymmetric magnetic order. In absence of well defined physical properties showing direct experimental evidences of a toroidal moment in magnetic systems, one of the important issues is to find a material in which specific magnetoelectric effects\cite{ISI:7} would reflect indirectly the influence of the toroidic moment. Here we analyze theoretically the magnetoelectric effects disclosed in the multiferroic phase of Ba$_2$CoGe$_2$O$_7$ (BCG)\cite{ISI:8, ISI:9} and show that due to its specific magnetic symmetry, which allows existence of spontaneous magnetization, polarization and toroidal moment as well as linear and non linear magnetoelectric effects, the existence of a spontaneous toroidal moment gives rise to spontaneous and field-induced toroidic effects. At the microscopic level the toroidic contribution to the electric polarization is shown to result from single-ion effects.\\ 
\indent The  $P\bar{4}2_1m1'$ paramagnetic space group of BCG\cite{ISI:10} has at the centre of the tetragonal Brillouin-zone ($\vec{k}$=0) five irreducible representations (IR’s) denoted $\tau_1-\tau_5$.\cite{ISI:11} The one-dimensional IR`s $\tau_1-\tau_4$ induce non-polar magnetic symmetries $P\bar{4}2_1m$, $P\bar{4}2_1m'$, $P\bar{4'}2_1m$ and $P\bar{4'}2_1m'$,
and are not associated with the transition to the ferroelectric phase observed in BCG below $T_N$ =6.7K. The 2-dimensional IR $\tau_5$ spanned by the order-parameter components $\eta _1=\rho cos(\theta)$ and $\eta _2=\rho sin(\theta)$ is associated with the Landau expansion:
\begin{equation}
F = \frac{\alpha}{2} \rho^2 + \frac{\beta_1}{4} \rho^4 + \frac{\beta_2}{4} \rho^4 cos(4\theta)+...+\frac{\gamma}{8}\rho^8 cos^2(4\theta)
\label{eq:F1}
\end{equation}
Minimizing \textit{F} yields 3 possible magnetically ordered phases below the paramagnetic phase: \\ 
1) Phase I corresponds to $\rho^e=\pm\left(\frac{\alpha_0(T_N-T)}{\beta_1+\beta_2}\right)^{\frac{1}{2}}$ and $cos(4\theta)=1$ ($\theta$=n$\frac{\pi}{2}$). It has the orthorhombic magnetic space groups $P2_1'2_12'$ ($\eta_1=\rho^e$,$\eta_2=0$) or $P2_12_1'2'$ ($\eta_1=0$,$\eta_2=\rho^e$), which form energetically equivalent domains of the same equilibrium phase.\\  
2) Phase II has the magnetic symmetry $Cm'a2'$ with $\rho^e=\pm\left(\frac{\alpha_0(T_N-T)}{\beta_1-\beta_2}\right)^{\frac{1}{2}}$  and $cos(4\theta)=-1$ ($\theta=(2n+1)\frac{\pi}{4}$), or equivalently $\eta_1=\pm\eta_2=\rho^e$. The phase involves a variety of spontaneous physical properties and domains represented in Fig. \ref{fig:F1}. Denoting $\vec{s_1}$ and $\vec{s_2}$ the magnetic spins associated with the Co$^{2+}$ ions located at (0,0,0) and (0.5,0.5,0) positions,\cite{ISI:10} the phase displays four weak ferromagnetic/antiferromagnetic domains with a spontaneous unit cell magnetization $\vec{M}=\vec{s_1}+\vec{s_2}$  along the $m'$-plane and unit cell antiferromagnetic vector $\vec{L}=\vec{s_1}-\vec{s_2}$ along the $a$-plane. The polar symmetry of the phase also gives rise to two ferroelectric ($\pm P_z$) - ferroelastic ($\pm e_{xy}$)  domains, and four \textit{toroidic domains} corresponding to the spontaneous toroidal moment $\vec{T}=\hat{\nu }(\vec{M}\times \vec{P})$, collinear to $\vec{L}$, where $\hat{\nu}$ is a third rank tensor.\\
3) Phase III of symmetry $P2'$ is stabilized for $\theta\neq n \frac{\pi}{4}$ and requires an eighth-degree expansion of $F$, involving 8 weak-ferromagnetic/antiferromagnetic domains and 2 ferroelectric-ferroelastic domains.\\
\indent The theoretical phase diagram containing the preceding phases is shown in Fig. \ref{fig:F2}. Phases II and III allow a spontaneous polarization along $z$, as observed experimentally in BCG\cite{ISI:8,ISI:9}. However, only phase II can be reached directly from the paramagnetic phase whereas a transition to phase III goes across phases I or II, or displays a first-order character. The temperature dependence of the polarization varies continuously at T$_N$, and the configuration of the spin moments\cite{ISI:9} is consistent with the four domains of phase II. Therefore, phase II can unambiguously be identified as the ferroelectric phase arising below T$_N$. In this phase $F$ can be written in function of $\vec{M}=(M_x,M_y), \vec{L}=(L_x,L_y)$  and  $P_z$  as:
\begin{eqnarray}
F = a_1L^2 + a_2L^4 + b_1M^2+b_2M^4+ \nonumber \\ 
c(M_xL_x-M_yL_y) + \frac{P^2_z}{2\varepsilon^0_{zz}} + \delta_1L_xL_yP_z + \nonumber \\
\delta_2M_xM_yP_z + \delta_3(M_xL_y-M_yL_x)P_z
\label{eq:F2}
\end{eqnarray}
where $\varepsilon^0_{zz}$ is the dielectric permittivity in the paramagnetic phase, and $a_i, b_i, c$ and $\delta_i$ are phenomenological coefficients. The equilibrium polarization below $T_N$ reads:
\begin{eqnarray}
P^e_z = -\varepsilon^0_{zz}\left[\delta_1L_xL_y+\delta_2M_xM_y + \delta_3(M_xL_y-M_yL_x)\right]
\label{eq:F3}
\end{eqnarray}
The two first terms into brackets express the respective contributions of the antiferromagnetic and weak-ferromagnetic order-parameters to the polarization, whereas the third term reflects their coupling contribution. Since $\left|\vec{L} \right|$ and $\left|\vec{M} \right|$ vary below T$_N$ as  $\sim(T_N-T)^{\frac{1}{2}}$, $P^e_z$ varies as $(T_N-T)$, consistent with the linear dependence observed for $P^e_z(T)$,\cite{ISI:8} corresponding  to an improper ferroelectric critical behaviour. The dielectric permittivity varies as  $\varepsilon_{zz}(T)=\varepsilon^0_{zz}\left( 1+\frac{\varepsilon^0_{zz}\delta ^2_1}{2a_2}\right)$  for $T<T_N$, in agreement with the reported upward discontinuity at $T_N$.\cite{ISI:8}\\
The  $\delta_3$-term in Eq. (\ref{eq:F2}) reflects the invariance of the mixed vector product $(\vec{M}\times\vec{L})\cdot \vec{P}$ under the symmetry operations of the paramagnetic phase. Analogously, the existence of the $(T_xM_y-T_yM_x)P_z$ invariant  involving the toroidal moment components  ($T_x,T_y$), expresses the invariance of the mixed vector products $(\vec{T}\times\vec{M})\cdot\vec{P}$  and $(\vec{T}\times\vec{P})\cdot\vec{M}$ which yield the following relationships between the spontaneous components:
\begin{equation}
\vec{P} = \hat{\mu}(\vec{T}\times\vec{M})
\label{eq:F4}
\end{equation}
and 
\begin{equation}
\vec{M} = \hat{\lambda }(\vec{T}\times\vec{P})
\label{eq:F5}
\end{equation}
where $\hat{\mu } $  and $\hat{\lambda } $ are third rank tensors. Applying electric or magnetic fields the existence of a spontaneous toroidal moment $\vec{T}^s $ gives:
\begin{equation}
\vec{P} = \vec{P}^s + \hat{\varepsilon}\vec{E} + \hat{\alpha}\vec{H}  + \hat{\sigma }^H(\vec{T}\times\vec{H})
\label{eq:F6}
\end{equation}
\begin{equation}
\vec{M} = \vec{M}^s + \hat{\chi }\vec{H} + \hat{\beta }\vec{E}  + \hat{\sigma }^E(\vec{T}\times\vec{E})
\label{eq:F7}
\end{equation}
\begin{equation}
\vec{T} = \vec{T}^s + \hat{\kappa}^E\vec{E} + \hat{\kappa}^M\vec{H}  + \hat{\sigma }^{EH}(\vec{E}\times\vec{H})
\label{eq:F8}
\end{equation}

where the third-rank tensors $\hat{\sigma}^H$ and $\hat{\sigma}^E$ precede additional polarization and magnetization  components induced by the coupling of the total toroidal moment $\vec{T} $ to non-collinear $\vec{H} $ or $\vec{E}$ fields. $\hat{\kappa}^E $ and $\hat{\kappa}^E $ are the electrotoroidal and magnetotoroidal tensors. The $\hat{\sigma}^{EH} $-term represents the induced toroidal contribution under non-collinear electric and magnetic fields. One should emphasize that the toroidal contributions to $\vec{P}$ and $\vec{M}$ in Eqs. (\ref{eq:F6}) and (\ref{eq:F7}), which are activated by a single external field non-collinear to $\vec{T}$, differ from the "ferromagnetotoroidic" and "ferroelectrotoroidic" effects\cite{ISI:12} implying the application of two distinct non-collinear fields. Although these contributions represent higher-order effects they can be differentiated  from the non-linear magnetoelectric contributions. For example, the non-linear $H^2$ contribution to $\vec{P}$ is observable above and below the transition while the $\vec{T}\times\vec{H}$ contribution is observable only below. More generally, the symmetries of the third-rank tensors $\hat{\sigma}^H$ and $\hat{\sigma}^E$ are different from the symmetries of the non-linear magnetic and electric susceptibilities, and the critical behaviour at constant fields of the corresponding tensor components is also different.\\
\indent Eq.(\ref{eq:F6}-to-(\ref{eq:F8})) provide an interpretation of the remarkable magnetoelectric  effects reported in BCG.\cite{ISI:8,ISI:9} Applying $H_z$ field gives rise to a polarization component $P_x$ increasing from 0 to $120\mu Cm^{-2}$   for $0<H_z<8T$. In the orthorhombic setting, which is turned by 45$^\circ$ with respect to the tetragonal axes (Fig. \ref{fig:F1}), one gets from Eq. (\ref{eq:F6}):
\begin{equation}
P_x(H_z) = \left[\alpha_{13}+T^s_y(\chi_{33}+\sigma^H_{123})\right]H_z 
\label{eq:F9}
\end{equation}
Consistent with the linear increase observed for $P_x(H_z)$ with increasing field\cite{ISI:8} and with its sign reversal on reversing $H_z$.\cite{ISI:9}  The sharp increase of $P_x(T)$  below $T_N$ at constant field,\cite{ISI:8} denotes a substantial toroidal contribution $\sigma^H_{123}T^s_y\approx(T_N-T)^{\frac{3}{2}}$, with respect to the linear magnetoelectric contribution $\alpha_{13}\approx(T_N-T)^{\frac{1}{2}}$. The observed increase of $P_z(H_z)$  from  $-11\mu Cm^{-2}$ at $H_z=0$, to $+80\mu Cm^{-2}$ in 8T is given by:
\begin{equation}
P_z(H_z)  = P^s_z+(\chi _{23}T_x - \chi _{13}T^s_y)H_z
\label{eq:F10}
\end{equation}
The shift of $P_z(H_z)$ to higher temperature under applied field\cite{ISI:8} is due to the renormalization of the coefficient $a_1 \approx (T-T_N)$ in Eq. (\ref{eq:F2}), which increases $T_N$ by $T_N(H_z)-T_N(0) \approx \chi _{33}H^2_z$. In order to account for the even dependence of $H_z$ observed for $P_z(H_z)$, one has to consider a higher order contribution, e.g. $\simeq H^2_z$, to $P_z(H_z)$.\\ 
\indent Other magnetoelectric effects have been reported\cite{ISI:9} under application of $H_{xy}$ and $H_{\bar{x}y}$ fields. $P_z$ \textit{increases} by increasing $H_{xy}$ and \textit{decreases} when increasing $H_{\bar{x}y}$.\cite{ISI:9} Projecting Eq. (\ref{eq:F6}) along $z$, one gets: 
\begin{eqnarray}
P_z(H_{xy})  = -\Delta P_z(H_{\bar{x}y})= \nonumber \\
\frac{1}{2}(\alpha_{31}+\sigma^H_{321}T^s_y+\sigma^H_{312}T^s_x)H_{xy}
\label{eq:F11}
\end{eqnarray}
Turning the $H_{xy}$ field by 90$^{\circ}$ transforms a ferroelectric domain into another, changing the sign of $P_z$.\cite{ISI:9} As for $P_z(H_z)$, a shifting of the transition temperature is observed under $H_{xy}$ field,\cite{ISI:9}  $P_z(T)$ decreasing smoothly down to $T_N$=12K for $H_{xy}$=5T , with a $(T_N-T)^{\frac{3}{2}}$ critical dependence of 
$A(T^s_x,T^s_y)=\sigma^H_{321}T^s_y+\sigma^H_{312}T^s_x$. These effects occur at low magnetic fields. At higher fields $P_x(H_{xy})$ decreases and changes sign.\cite{ISI:9} This behaviour, assumed to correspond to a spin-structural change,\cite{ISI:9} requires including the higher-order invariant $(\vec{T^s}\times \vec{H})\cdot (\vec{T^s}\cdot \vec{H}) \approx \frac{K(T^s_x,T^s_y)}{2}H^2_{xy}$ in Eq. (\ref{eq:F6}). For  $K<0$, $P_z(H_{xy})$ decreases above the threshold field $H^{th}_{xy}=-\frac{\alpha_{31}+A}{4K}$ taking negative values for $H_{xy}>2H^{th}_{xy}$.\\        
\indent To gain insight into the nature of the magnetic interactions governing the magnetoelectric and toroidic behaviours of BCG, let us express the order-parameter components in function of the magnetic spins $\vec{s_1}$ and $\vec{s_2} $. Writing $\vec{s_i}=s^a_i\vec{a}+s^b_i\vec{b}+s^c_i\vec{c}$ ($i=1,2$), where $\vec{a}, \vec{b}, \vec{c}$ are the tetragonal lattice vectors, the representation $\Gamma $ transforming the $s^{a,b,s}_i $ components decomposes into $\Gamma =\tau_1+\tau_2+2\tau_5$, i.e. \textit{two order-parameter copies}, denoted $(\eta_1,\eta_2)$ and $(\zeta _1,\zeta _2)$, \textit{are involved in the transition mechanism}.  Standard projector techniques\cite{ISI:13} give:
\begin{eqnarray}
\eta _1  = s^a_1+s^a_2, \eta _2  = -(s^b_1+s^b_2) \nonumber \\
\zeta  _1 = s^b_1-s^b_2,  \zeta  _2 = s^a_1-s^a_2
\label{eq:F12}
\end{eqnarray}
It shows that the two order-parameter copies coincide with the ferromagnetic and antiferromagnetic vectors. On the other hand, projections of $\Gamma $ on $\tau_1$ and $\tau_2$ lead to $s^c_1-s^c_2=0$ and $s^c_1+s^c_2=0$, i.e. $s^c_1=s^c_2=0$, confirming the in-plane spin ordering in BCG. The equilibrium values of $(\eta _1,\eta _2)$ and $(\zeta  _1,\zeta  _2)$ in phase II yield the spin configurations for the four magnetic domains represented in Fig 1, namely: two weak ferromagnetic domains for $s^a_1+s^a_2=\pm (s^b_1+s^b_2)$, and two antiferromagnetic domains for $s^b_1-s^b_2=\pm (s^a_1-s^a_2)$.  The spontaneous polarization $P^e_z$ at zero field reads:
\begin{eqnarray}
P^e_z  = \delta '_1 \eta_1\eta_2 + \delta '_2 \zeta _1\zeta _2 + \delta '_3 (\eta_1\zeta _2 + \eta_2 \zeta _1)
\label{eq:F13}
\end{eqnarray}

analogue to Eq. (3). Using Eq. (12) yields:
\begin{eqnarray}
P^e_z = \delta '_1 (s^a_1s^b_1+s^a_1s^b_2+s^a_2s^b_1+s^a_2s^b_2) \nonumber \\
+ \delta '_2 (s^a_1s^b_1-s^b_1s^a_2-s^b_2s^a_1+s^a_2s^b_2) \nonumber \\
+ \delta '_3 (s^{a2}_1-s^{a2}_2-s^{b2}_2+s^{b2}_2)
\label{eq:F14}
\end{eqnarray}
Eq. (\ref{eq:F14}) holds for a pair of antiferromagnetic domains (e.g. $\eta^e_1=\eta^e_2$  and $\zeta ^e_1=\zeta ^e_2$) whereas $-P^e_z$ coincides with the other pair ($\eta^e_1=-\eta^e_2, \zeta ^e_1=-\zeta ^e_2$). The $\delta '_1$, $\delta '_2$ and $\delta '_3$ terms represent the respective contributions of the spontaneous ferromagnetic, antiferromagnetic \textit{and toroidal} contributions to the spontaneous polarization arising in the multiferroic state. The $\delta '_3$ term, which is the microscopic analogue of the spontaneous toroidic effect given by Eq. (\ref{eq:F4}), reflects \textit{single-ion effects}, while the two other terms contain invariants $s^u_is^{\nu }_i (i=1,2; u,\nu =a,b)$, also corresponding to single-ion effects, and $s^u_is^{\nu }_j (i \neq j)$ invariants expressing the symmetric part of the exchange coupling interaction between the two Co spins. These results support the interpretation\cite{ISI:8} that the spin-dependent $p-d$ hybridization between the transition-metal (Co) and ligand (O) contributes to the ferroelectricity in BCG via the spin-orbit interaction, as well as the proposed mechanism of lattice relaxation induced by exchange striction.\cite{ISI:9} Note that the Dzialoshinskii-Moriya (DM) interaction does not contribute directly to the polarization but is responsible of the canting inducing the weak-ferromagnetic moments,\cite{ISI:10} which stabilizes the toroidal moment giving rise to the $\delta '_3$  term in Eq. (\ref{eq:F14}).\\
\indent In conclusion, our theoretical analysis shows that in multiferroic compound exhibiting spontaneous magnetization, polarization and toroidal moment, field-induced toroidic effects occur, consisting of additional toroidal contributions to the polarization and magnetization. These toroidal effects allow a comprehensive description of the specific magnetoelectric properties observed in BCG, which have been shown to relate to additional toroidal contributions to the polarization, corresponding at the microscopic level to single site magnetic interactions. They should allow clarifying the intrinsic role of the toroidal moment in magnetic multiferroics.


\begin{thebibliography}{14}
\expandafter\ifx\csname natexlab\endcsname\relax\def\natexlab#1{#1}\fi
\expandafter\ifx\csname bibnamefont\endcsname\relax
  \def\bibnamefont#1{#1}\fi
\expandafter\ifx\csname bibfnamefont\endcsname\relax
  \def\bibfnamefont#1{#1}\fi
\expandafter\ifx\csname citenamefont\endcsname\relax
  \def\citenamefont#1{#1}\fi
\expandafter\ifx\csname url\endcsname\relax
  \def\url#1{\texttt{#1}}\fi
\expandafter\ifx\csname urlprefix\endcsname\relax\def\urlprefix{URL }\fi
\providecommand{\bibinfo}[2]{#2}
\providecommand{\eprint}[2][]{\url{#2}}

\bibitem[{\citenamefont{Fiebig}({2005})\citenamefont{Fiebig}}]{ISI:1}
\bibinfo{author}{\bibfnamefont{M.}~\bibnamefont{Fiebig}},
  \bibinfo{journal}{{J. Phys. D }} \textbf{\bibinfo{volume}{{38}}},
  \bibinfo{pages}{{R123}} (\bibinfo{year}{{2005}}).
  
\bibitem[{\citenamefont{Spaldin et~al.}({2008})\citenamefont{Spaldin, Fiebig, and Mostovoy}}]{ISI:2}
\bibinfo{author}{\bibfnamefont{N.~A.}~\bibnamefont{Spaldin}},
  \bibinfo{author}{\bibfnamefont{M.}~\bibnamefont{Fiebig}}, \bibnamefont{and}
  \bibinfo{author}{\bibfnamefont{M.}~\bibnamefont{Mostovoy}},
  \bibinfo{journal}{{J. Phys.: Condens. Matter }} \textbf{\bibinfo{volume}{{20}}},
  \bibinfo{pages}{{434203}} (\bibinfo{year}{{2008}}).

\bibitem[{\citenamefont{Kopaev}({2009})\citenamefont{Kopaev}}]{ISI:3}
\bibinfo{author}{\bibfnamefont{Yu.~V.}~\bibnamefont{Kopaev}},
  \bibinfo{journal}{{Physics Uspekhi}} \textbf{\bibinfo{volume}{{52}}},
  \bibinfo{pages}{{1111}} (\bibinfo{year}{{2009}}).
  
\bibitem[{\citenamefont{Dubovik and Tugushev}({1990})\citenamefont{Dubovik and Tugushev}}]{ISI:4}
\bibinfo{author}{\bibfnamefont{V.~M.}~\bibnamefont{Dubovik}}, \bibnamefont{and}
  \bibinfo{author}{\bibfnamefont{V.~V.}~\bibnamefont{Tugushev}}, 
  \bibinfo{journal}{{Phys. Rep.}} \textbf{\bibinfo{volume}{{187}}},
  \bibinfo{pages}{{145}} (\bibinfo{year}{{1990}}).
  
\bibitem[{\citenamefont{Gorbatsevitch and Kopaev}({1994})\citenamefont{Gorbatsevitch and Kopaev}}]{ISI:5}
\bibinfo{author}{\bibfnamefont{A.~A.}~\bibnamefont{Gorbatsevitch}},
\bibinfo{author}{\bibfnamefont{Yu.~V.}~\bibnamefont{Kopaev}},
  \bibinfo{journal}{{Ferroelectrics}} \textbf{\bibinfo{volume}{{161}}},
  \bibinfo{pages}{{321}} (\bibinfo{year}{{1994}}).  
  
\bibitem[{\citenamefont{Van Aken et~al.}({2007})\citenamefont{Van Aken, Rivera, Schmid, and Fiebig}}]{ISI:6}
\bibinfo{author}{\bibfnamefont{B.~B.}~\bibnamefont{Van Aken}},
  \bibinfo{author}{\bibfnamefont{H.~P.}~\bibnamefont{Rivera}},
  \bibinfo{author}{\bibfnamefont{H.}~\bibnamefont{Schmid}}, \bibnamefont{and}
  \bibinfo{author}{\bibfnamefont{M.}~\bibnamefont{Fiebig}},  
  \bibinfo{journal}{{Nature (London)}} \textbf{\bibinfo{volume}{{449}}},
  \bibinfo{pages}{{702}} (\bibinfo{year}{{2007}}).  

\bibitem[{\citenamefont{Mettout, Toledano and Fiebig}({2010})\citenamefont{Mettout, Toledano, and Fiebig}}]{ISI:7}
\bibinfo{author}{\bibfnamefont{B.}~\bibnamefont{Mettout}},
  \bibinfo{author}{\bibfnamefont{P.}~\bibnamefont{Toledano}}, \bibnamefont{and}
  \bibinfo{author}{\bibfnamefont{M.}~\bibnamefont{Fiebig}},
  \bibinfo{journal}{{Phys. Rev. B}} \textbf{\bibinfo{volume}{{81}}},
  \bibinfo{pages}{{214417}} (\bibinfo{year}{{2010}}).
  
\bibitem[{\citenamefont{Yi et~al.}({2008})\citenamefont{Yi, Choi, Lee and Cheong}}]{ISI:8}
\bibinfo{author}{\bibfnamefont{H.~T.}~\bibnamefont{Yi}},
  \bibinfo{author}{\bibfnamefont{Y.~J.} \bibnamefont{Choi}},
  \bibinfo{author}{\bibfnamefont{S.} \bibnamefont{Lee}},
  \bibinfo{author}{\bibfnamefont{S.~W.}~\bibnamefont{Cheong}}, \bibnamefont{and}
  \bibinfo{journal}{{Appl. Phys. Lett.}}
  \textbf{\bibinfo{volume}{{92}}}, \bibinfo{pages}{{212904}}
  (\bibinfo{year}{{2008}}). 

\bibitem[{\citenamefont{Murakawa et~al.}({2010})\citenamefont{H. Murakawa, Y. Onose, S. Miyahara, N. Furukawa, and Y. Tokura}}]{ISI:9}
\bibinfo{author}{\bibfnamefont{H.}~\bibnamefont{Murakawa}},
  \bibinfo{author}{\bibfnamefont{Y.}~\bibnamefont{Onose}},
  \bibinfo{author}{\bibfnamefont{S.}~\bibnamefont{Miyahara}},
  \bibinfo{author}{\bibfnamefont{N.}~\bibnamefont{Furukawa}}, \bibnamefont{and}
  \bibinfo{author}{\bibfnamefont{Y.}~\bibnamefont{Tokura}}, 
  \bibinfo{journal}{{Phys. Rev. Lett. }} \textbf{\bibinfo{volume}{{105}}},
  \bibinfo{pages}{{137202}} (\bibinfo{year}{{2010}}).
  
 \bibitem[{\citenamefont{Zheludev et~al.}({2003})\citenamefont{Zheludev, Sato, Masuda, Uchinokura, Shirane, and Roessli}}]{ISI:10}
\bibinfo{author}{\bibfnamefont{A.}~\bibnamefont{Zheludev}},
  \bibinfo{author}{\bibfnamefont{T.}~\bibnamefont{Sato}},
  \bibinfo{author}{\bibfnamefont{T.}~\bibnamefont{Masuda}},
  \bibinfo{author}{\bibfnamefont{K.}~\bibnamefont{Uchinokura}}, \bibnamefont{and}
  \bibinfo{author}{\bibfnamefont{G.}~\bibnamefont{Shirane}}, 
  \bibinfo{author}{\bibfnamefont{B.}~\bibnamefont{Roessli}},  
  \bibinfo{journal}{{Phys. Rev. B}} \textbf{\bibinfo{volume}{{68}}},
  \bibinfo{pages}{{024428}} (\bibinfo{year}{{2003}}). 
  
 \bibitem[{\citenamefont{Kovalev}(1965)}]{ISI:11}
\bibinfo{author}{\bibfnamefont{O.~V.} \bibnamefont{Kovalev}},
  \emph{\bibinfo{title}{The irreducible representations of Space Groups}}
  (\bibinfo{publisher}{Gordon an Breach},
  \bibinfo{address}{New York}, \bibinfo{year}{1965}). 
   
\bibitem[{\citenamefont{Schmid}({2003})\citenamefont{Schmid}}]{ISI:12}
\bibinfo{author}{\bibfnamefont{H.}~\bibnamefont{Schmid}},
  \bibinfo{journal}{{J. Phys. : Condens. Matter}} \textbf{\bibinfo{volume}{{20}}},
  \bibinfo{pages}{{434201}} (\bibinfo{year}{{2008}}).  
  
\bibitem[{\citenamefont{Toledano and Toledano}(1987)}]{ISI:13}
\bibinfo{author}{\bibfnamefont{J.~C.} \bibnamefont{Toledano}}, \bibnamefont{and}
\bibinfo{author}{\bibfnamefont{P.} \bibnamefont{Toledano}},
  \emph{\bibinfo{title}{The Landau Theory of Phase Transitions}}
  (\bibinfo{publisher}{World Scientific},
  \bibinfo{address}{ Singapore}, \bibinfo{year}{1987}).    
  
\end{thebibliography}
\end{document}